\documentclass[showkeys,twocolumn,showpacs,preprintnumbers,amsmath,amssymb,prl]{revtex4}
\usepackage{Depken,graphicx,psfrag}

\begin{document}
\title{Fluctuation-dissipation relation and the Edwards entropy for a glassy granular compaction model}
\author{Martin Depken and Robin Stinchcombe}
 \affiliation{University of Oxford, Department of Physics, Theoretical Physics\\ 1 Keble Road, Oxford, OX1 3NP, U.K.}
 \altaffiliation[MD's present address: ]{Instituut-Lorentz, Leiden University, P.O. Box 9506, 2300 RA Leiden, Netherlands\\ email: depken@lorentz.leidenuniv.nl}
\begin{abstract}
We analytically study a one dimensional compaction model in the glassy regime. Both correlation and response functions are calculated exactly in the evolving dense and low tapping strength limit, where the density relaxes in a $1/\ln t$ fashion. The response and correlation functions turn out to be connected through a non-equilibrium generalisation of the fluctuation-dissipation theorem. The initial response in the average density to an increase in the tapping strength is shown to be negative, while on longer timescales it is shown to be positive. On short time scales the fluctuation-dissipation theorem governs the relation between correlation and response, and we show that such a relationship also exists for the slow degrees of freedom, albeit with a different temperature. The model is further studied within the statistical theory proposed by Edwards and co-workers, and the Edwards entropy is calculated in the large system limit. The fluctuations described by this approach turn out to match the fluctuations as calculated through the dynamical consideration. We believe this to be the first time these ideas have been analytically confirmed in a non-mean-field model.
\end{abstract}
\maketitle
Granular materials have had much experimental and theoretical attention in recent years. They are intriguing as they form an additional state of matter, fundamentally different from gases, liquids and solids~\cite{Jaeger96}.  Specific for these systems is that the thermal energy of its constituents, i.e. the grains, is negligible compared to other relevant energy scales. Since the thermal energy is negligible, there is no inherent mechanism that makes a granular system explore its phase space. Any energy fed to such system is  quickly dissipated, and if left unperturbed the system becomes trapped in one of many metastable states with an essentially infinite lifetime. In many situations where these materials are handled or used in production, they are continuously fed energy through external perturbations. As a result the system starts to explore the available phase space and macroscopic quantities, such as the density, start to evolve.  This situation has been experimentally examined~\cite{Knight95,Nowak98} through subjecting a container filled with a granular material to many  taps of acceleration $\Gamma$. The time between taps was large enough for the system to dissipate any excess energy, and settle into one of its many meta-stable states between each tap. The relaxation of density (or free-volume fraction) in these experiments was well fitted by an inverse logarithmic form. In a different context we have  previously~\cite{Stinchcombe02,Depken03} introduced a simple one dimensional model for which we were able to analytically derive this inverse logarithmic relaxation. We here consider a robust generalization of the model, and obtain both response and correlation functions through a direct dynamical approach.  The system is further investigated in the context of Edwards' statistical theory of granular compaction~\cite{Edwards89,Mehta89,Oakshot92}, and the result of the two approaches are compared.

\begin{figure}[htp]
\begin{center}
  \psfrag{ex}{\small $r_{\rm e}(\Gamma)$}
  \psfrag{D}[]{\small $D(\Gamma)$}
  \psfrag{L}{\small $L$}
  \psfrag{1}[]{\small $r_{\rm d}$}
  \includegraphics[width=0.9\columnwidth]{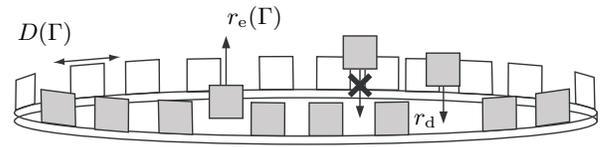}
  \caption{\label{fig:model}Schematic representation of the model with dynamical rules indicated.}
\end{center}
\end{figure}

The model can be seen as a minimal model of the bottom layer in a granular material compacting under tapping. It is related to the continuum car-parking model~\cite{Kolan99} and consists of unit sized hard-core blocks (particles) positioned on a ring of length $L$. The blocks, which interact via hard-core repulsion, do a caged diffusion along the ring with diffusion constant $D(\Gamma)$. The blocks are further able to evaporate from the ring at the rate $r_{\rm e}(\Gamma)=\exp(-f(\Gamma))$. To model tapping induced diffusion and activated escape from the ring one could for example choose $D(\Gamma)\propto\Gamma$ and $f(\Gamma)\propto 1/\Gamma$. When a gap of size larger than one opens up, we take it to be filled by a random deposition of a particle with the tapping-strength-independent rate $r_{\rm d}=\O(\Gamma^0)>0$. This is meant to reflect the fact that the gravitational pull on the particles is independent of the tapping strength. These rules are summarised in Figure~\ref{fig:model}. 
Since we are interested in the very slow dynamics exhibited by this model in the limit of a dense system subject to a weak tapping, we consider the weak tapping strength limit, in which we demand (consistent with diffusion and activated evaporation)
\be{eq:ord}
  r_{\rm e}(\Gamma), D(\Gamma)/\epsilon^2\ll r_{\rm d}, \quad \Gamma\searrow 0,
\ee
where $\epsilon$ is the free length fraction. It is this ordering of the rates, and not their actual form, that plays a crucial role in determining the long time dynamics. The average time a gap of size larger than one, stays larger than one, before it is closed by diffusion, is proportional to $1/D(\Gamma)$. The time it takes to fill such a gap by a deposition event is proportional to $1/r_{\rm d}$. Due to the above ordering of rates we see that in the weak tapping limit all gaps that open up will eventually be filled by a deposition. Therefore the effect on the long time dynamics of any evaporation events is suppressed. Thus we can use the effective rules 
$$
  r_{\rm e}= 0, \quad r_{\rm d}=\infty, \quad D(\Gamma)={\rm finite,}
$$
for the long time evolution of the system. This shows that as long as the different rates satisfy~(\ref{eq:ord}), then the long time dynamics is insensitive to the precise form of the evaporation and deposition rates. Taking this limit amounts to completely suppressing any fast processes, such as the evaporation of a block followed by a subsequent deposition of a block in the created gap. 

\begin{figure}
  \psfrag{Ref}{\small reflecting boundary}
  \psfrag{A}{\small $\Delta_1$}
  \psfrag{B}{\small $\Delta_2$}
  \psfrag{C}{\small $\Delta_3$}
  \psfrag{0}{\small $0$}
  \psfrag{1}{\small $1$}
  \psfrag{Port}{\small portal: $\pi_3\rightarrow\pi_4$}
  \includegraphics[width=0.9\columnwidth]{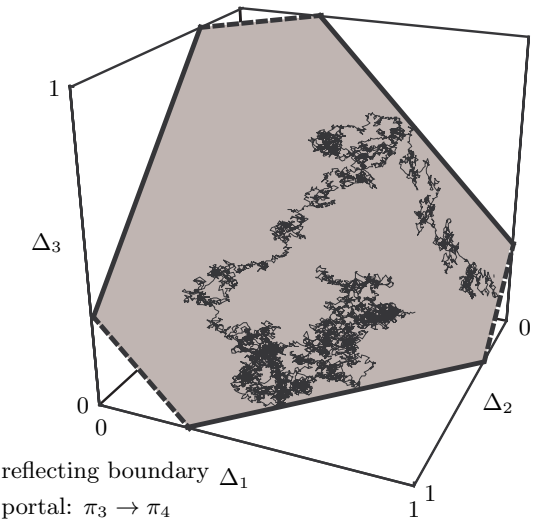}
  \caption{\label{fig:pi} Diffusion on the hyperplane showing a few reflections at the boundary, corresponding to two blocks bouncing off each other, and an eventual escape and transfer to the hyper plane $\pi_4$ through a gap of size 1 opening up and then being filled with a block.}
\end{figure}

\begin{figure}
\psfrag{2}{}
\psfrag{1}[c]{$1$}
\psfrag{a}[c]{$10^{8}$}
\psfrag{b}[c]{$10^{16}$}
\psfrag{c}{$\chi_\Gamma(t,t_{\rm w})$}
\psfrag{t}{$t-t_{\rm w}$}
\includegraphics[width=0.9\columnwidth, height=0.3\columnwidth]{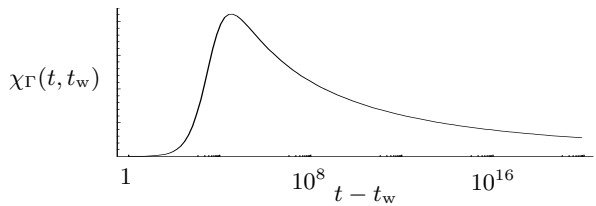}
\caption{\label{fig:response} The form (scale arbitrary) of the response function of the slow degrees of freedom, $\chi_\Gamma(t,t_{\rm w})$, as a function of $t-t_{\rm w}$, and with a initial free-length fraction at $t_{\rm w}$, $\epsilon_{\rm w}=0.05$. Similar response functions have previously been found in~\cite{Barrat00,Nicodemi99}.}
\end{figure}

In a different context we have previously~\cite{Stinchcombe02,Depken03} derived the exact form of the density-density correlation function in the dense and low tapping strength limit for the specific choice of $D(\Gamma)\propto \Gamma$ and $f(\Gamma)\propto 1/\Gamma$. This was done through a geometrical description of the problem, and we now extend the considerations to the present case, and include a calculation of the response function of the slow degrees of freedom. We view the time evolution of the system between deposition events as a diffusion of the gaps (between blocks) on the hyper surface of constant density
\be{eq:pi}
\pi_N=\left\{\bar\Delta_{N}\left| \sum_{n=1}^{N}\Delta_n=L-N, \,\,0\le \Delta_n <1\right.\right\},
\ee
where $\bar\Delta_N$  is a vector consisting of all the $N$ individual gap sizes between $N$ adjacent blocks. This is illustrated in Figure~\ref{fig:pi} for the case of a ring with only three blocks.
Asymptotic (large system, low tapping strength, high density) dynamic properties of the system can now be calculated through assuming ergodicity on $\pi_N$ between deposition events, and using straightforward geometrical considerations. Ergodicity between deposition events should hold in the dense limit since the time between deposition events diverges, and thus the diffusive motion has enough time to relax the system between these events. For further technical details we refer to~\cite{Stinchcombe02,Depken03,Depken04}. (The intermediate time regime, in which the system ought to display spatial structures, is at present being investigated~\cite{Stinchcombe04}).  The connected two time density-density correlation function has the form
$$
  C^{\rm c}(t,t_{\rm w})=\frac{T_2(\epsilon_{\rm w},\epsilon_0)}{L\tau(\epsilon)\tau(\epsilon_{\rm w})}\sim \frac{\epsilon_{\rm w}^4}{2L\epsilon^2}e^{1/\epsilon_{\rm w}-1/\epsilon},
$$
with $\epsilon=\epsilon(t|\epsilon_0,t_0)$ and $\epsilon_{\rm w}=\epsilon(t_{\rm w}|\epsilon_0,t_0)$, where $\epsilon(t|\epsilon_0,t_0)$ denotes the average free-length fraction on the ring at time $t$, given the initial free-length fraction $\epsilon_0$ at time $t_0$. In the above $t_{\rm w}$ is the waiting time, and $\tau(\epsilon)\sim k \epsilon^2 e^{1/\epsilon}/D(\Gamma)$ is the average time between deposition events within a unit length of the ring, given the free-length fraction $\epsilon$. Here $k$ is a constant only dependent on the geometry of the hyper planes. We have further used
\begin{multline*}
  T_n(\epsilon_{\rm w},\epsilon_0)=\int^{\epsilon_0}_{\epsilon_{\rm w}} \d \epsilon\, \tau^n(\epsilon)\\\sim\left(\frac{k}{D(\Gamma)}\right)^n \frac{1}{n}\left(\epsilon_{\rm w}^{2(n+1)}(1+\O(\epsilon_{\rm w}))e^{n/\epsilon_{\rm w}}\right.\\
-\left. \epsilon_{0}^{2(n+1)}(1+\O(\epsilon_0))e^{n/\epsilon_0}\right).
\end{multline*}
Thus we see that the long time density-density correlation is independent not only of the evaporation and deposition rates, but also of the diffusion constant. Since $T_1(\epsilon,\epsilon_0)$ by definition is the average time it takes the system to evolve from a free-length fraction $\epsilon_0$ to a free-length fraction $\epsilon$, we have an implicit relationship for the evolution of the free-length fraction 
$$
T_1(\epsilon(t|\epsilon_0,t_0),\epsilon_0)=t-t_0.
$$
Through differentiating the above with respect to $\Gamma$ it is an easy matter to calculate the response in density to a change of the tapping strength starting at time $t_{\rm w}$:
\begin{multline*}
  \chi_\Gamma(t,t_{\rm w})=-\frac{\partial \epsilon(t|\epsilon_{\rm w},t_{\rm w})}{\partial \Gamma}=\frac{D'(\Gamma)T_1(\epsilon,\epsilon_{\rm w})}{D(\Gamma)\tau(\epsilon)}\\
  \sim\frac{\epsilon^2 D'(\Gamma)}{D(\Gamma)}\lp 1-\lp\frac{\epsilon_{\rm w}}{\epsilon}\rp^4\exp(1/\epsilon_{\rm w}-1/\epsilon)\rp.
\end{multline*}
In Figure~\ref{fig:response} we display the response as a function of time.
 Since we have both response and correlation functions we are in position to consider a possible extension of the equilibrium fluctuation-dissipation theorem to the non-equilibrium situation present in our model. A direct comparison of the asymptotic form for the response and connected correlation functions gives
$$
  \chi_\Gamma(t,t_{\rm w})\sim \frac{2L D'(\Gamma)}{D(\Gamma)}\lp C^{\rm c}(t,t)-C^{\rm c}(t,t_{\rm w })\rp.
$$
In an equilibrium system the fluctuation-dissipation theorem states that the linear response and the correlation function (for intensive quantities) are related through
$$
  -\frac{\partial \chi_h(t,t_{\rm w})}{\partial t_{\rm w}}=\frac{V}{T_{\rm eq}}\frac{\partial C^{\rm c}(t,t_{\rm w})}{\partial t_{\rm w}}, 
$$
where $V$ is the system size and $h$ is the variable conjugate to the quantity considered. In our case the corresponding statement is
$$
 -\frac{\partial\chi_h(t,t_{\rm w})}{\partial t_{\rm w}}\sim\frac{L}{T_{\rm neq}}\frac{\partial C^{\rm c}(t,t_{\rm w})}{\partial t_{\rm w}}, \quad T_{\rm neq}=\frac{h'(\Gamma)D(\Gamma)}{2D'(\Gamma)},
$$
where $h(\Gamma)$ is the (unknown) variable conjugate to the density. It should be noted that in our asymptotic analysis we have nowhere demanded that $t\gg t_{\rm w}$, but just that $t-t_{\rm w}$ is larger than the relaxation time of the fast degrees of freedom, $t_{\rm fast}$. Thus the above results are valid as long as $t_{\rm w}\gg t_0$ and $t-t_{\rm w}\gg t_{\rm fast}$. Therefore $T_{\rm neq}$ could in general depend on some finite combination of $t$ and $t_{\rm w}$ in the limit $t_{\rm w}\rightarrow \infty$. It is thus striking that $T_{\rm neq}$ is completely independent of $t$ and $t_{\rm w}$. The temperature of the fast degrees of freedom can be directly calculated~\cite{Depken04} through the fluctuation-dissipation theorem as $T_{\rm eq}=h'/f'$, and therefore $T_{\rm eq}/ T_{\rm neq}=2D'/(D f')$. Thus, depending on which timescales we are considering, we  see different dynamically defined temperatures. This very behavior has previously been identified in mean-field models~\cite{Cugliandolo94} and through numerical simulations~\cite{Parisi97,Barrat99,Barrat00b}, and now we see it analytically in a non-mean-field model. For any reasonable system we have $D'>0$ and $f'<0$, and thus the two temperatures differ in sign (which has not been seen in mean field models). This sign difference arises because on the short timescale an increase in tapping strength decompactifies the system through the fast degrees of freedom considered above. In the aging regime though, a higher tapping strength increases the compactification rate. This behavior has previously been seen in systems with activated dynamics, such as the trap model~(see \cite{Sollich02} and references therein). We will later compare these results with those from considering the Edwards entropy of the system, which we derive next.

We will use the simplest form of this theory and apply it to our model, and for the omitted technical details we refer to~\cite{Depken03,Depken04}. A similar treatment has recently been carried out for the parking-lot model~\cite{Tarjus04}, and similar results for the simpler case when one allows gaps of size larger then one has been known for a long time~\cite{Tonks36}. The counting of the number of {\it blocked} configurations is central to the Edwards approach, and in our case these correspond to configurations for which no gap is larger than one. A powder driven by well separated periodic taps will explore the phase space of meta-stable or blocked states, consistent with the external conditions. It is then natural to define the entropy density in the thermodynamic limit as
$$
  s_{\rm Edw}(\epsilon)=\lim\limits_{L\rightarrow\infty}\frac{1}{L}\ln W_L(\epsilon),
$$
where $W_L(\epsilon)$ is the number of blocked microscopic states consistent with the free-length fraction $\epsilon$. We will refer to $s_{\rm Edw}$ as the Edwards entropy density. In analogy with equilibrium statistical mechanics one assumes that any of the states consistent with the macroscopic constraints $\epsilon$ and $L$ are equally probable. With this crucial assumption the statistical properties of the system are given by the micro-canonical partition function
$$
  Z(\epsilon,L)=\exp(L s_{\rm Edw}(\epsilon)).
$$
If we consider our original system as being part of a larger ensemble that allows exchange of particles between its subsystems, then we can move over to the canonical ensemble. We define the canonical partition function for the free-length fraction by
$$
  \Omega(\mu,L)=\int_0^1 \d \epsilon \exp(-L(\mu\epsilon-s_{\rm Edw}(\epsilon)),
$$
where $\mu$ is a Lagrangian multiplier ensuring the correct overall free length. In the thermodynamic limit we can use the saddle-point method to introduce the thermodynamic potential
$$
  f(\mu)=-\lim\limits_{L\rightarrow\infty}\frac{1}{L}\ln \Omega(\mu,L)=\mu \epsilon(\mu)-s_{\rm Edw}(\epsilon(\mu)) 
$$
where $\epsilon(\mu)$ solves $\partial_\epsilon s_{\rm Edw}(\epsilon(\mu))=\mu$. In the usual manner we can now calculate the average free-length fraction as a function of $\mu$, $\lv\epsilon\rv_\mu=\partial_\mu f(\mu)=\epsilon(\mu)$. Through the assumption that all possible states have {\it a priori} equal probability we have thus, with respect to the calculation of the steady state, been able to cut out any reference to the dynamics and replaced it with a statistical description using the thermodynamic variable $\mu$ conjugate to the free-length fraction. In our present setting the {\it a priori} equal probability is just the earlier used assumption of ergodicity on $\pi_N$, which should hold true in the dense limit. The Edwards entropy for the compaction model considered can then simply be expressed as
$$
  s_{\rm Edw}(\epsilon)=\lim\limits_{L\rightarrow \infty}\frac{1}{L} \ln {\rm Vol}(\pi_{L(1-\epsilon)}).
$$ 
where ${\rm Vol} (\pi_{L(1-\epsilon)})$ is the volume of the hyper surface $\pi_{L(1-\epsilon)}$ defined in~(\ref{eq:pi}). Asymptotic analysis~\cite{Depken03,Depken04} gives
\be{eq:implicit}
  \begin{array}{rl}
   s_{\rm Edw}(\epsilon)&=-g(\epsilon,s(\epsilon))\\
   g(\epsilon,s)&=2\epsilon s-(1-\epsilon)(\ln\sinh s-\ln s+s),
  \end{array}
\ee
where $s(\epsilon)$ solves $\partial_s g(\epsilon,s(\epsilon))=0$. Moving over to the canonical ensemble we need to solve $\mu=\partial_\epsilon s_{\rm Edw}(\epsilon(\mu))$ for the average free-length fraction $\lv\epsilon\rv_{\mu}=\epsilon(\mu)$ as a function of the thermodynamic variable $\mu$.  In the low tapping strength, high density, and large system limit this can be calculated as
$$  
  s_{\rm Edw}(\epsilon)\sim (1-\epsilon)\lp1-\ln\lp\epsilon^{-1}-1\rp\rp
$$
and thus $\epsilon(\mu)=\mu^{-1}(1+\O(\mu^{-1}\ln\mu))$. The density fluctuations can be written as
$$
  \lv \delta\epsilon^2\rv_{{\rm Edw}}= -\frac{1}{L}\partial_\mu \epsilon(\mu)\sim\frac{1}{L\mu^2}\sim\frac{\epsilon^2(\mu)}{L},
$$
which should be compared with the fluctuations of the slow degrees of freedom,  $C^{\rm c}(t,t)\sim \epsilon^2/2L$, as given by the dynamical considerations above. Though different in origin they agree in their functional dependence on $\epsilon$. Results with similar implications have previously been reported in numerical simulations of compaction models~\cite{Sellitto98, Barrat00b, Barrat01}. 

Lastly we note that in a steady state the free-length fraction, $\epsilon(\Gamma)$, is set by the condition that the effective evaporation rate for a typical configuration matches the deposition rate into gaps opened by diffusion,
$$
  \exp(-(f(\Gamma)+r_{\rm d}\epsilon^2(\Gamma)/D(\Gamma)))\sim \exp(-1/\epsilon(\Gamma)).
$$
For the natural choice $f(\Gamma)=1/k \Gamma$, and $D(\Gamma)\propto \Gamma$, this gives $\epsilon(\Gamma)\sim k\Gamma$, which compared to $\epsilon(\mu)$, as given by the Edwards picture, gives $\mu\sim1/k\Gamma$. Thus, in this setting it is natural to interpret $\Gamma$ as a temperature for the slow degrees of freedom.

In this letter we have presented what we believe is the first instance where an asymptotically exact calculation gives an extension of the fluctuation-dissipation theorem in the non-equilibrium regime of a granular compaction model. The results are very robust with respect to the different types of driving, with only the relation~(\ref{eq:ord}) setting the limits for the possible forms. Through the fluctuation-dissipation theorem it is possible to define two different temperatures, one for the fast and one for the slow degrees of freedom. Due to the difference in the response to a perturbation in the tapping strength of the short time and long time degrees of freedom, these temperatures have a different sign.  We have further treated the model with the Edwards theory of powders and calculated an exact (implicit) form of the Edwards entropy density. The fluctuations as calculated within the Edwards' picture further accurately describe the long-time dynamically induced fluctuations. 

The authors gratefully acknowledge Juan P. Garrahan and an anonymous referee for valuable comments. This work was supported by EPSRC under the Oxford Condensed Matter Theory Grant No. GR/R83712/01 and GR/M04426. MD gratefully acknowledges support from the Merton College Oxford Domus scholarship fund, and the Royal Swedish Academy of Science.  
\bibliography{bibliography}
\end{document}